\documentclass{jfm}
\usepackage{graphicx}
\usepackage{epstopdf, epsfig}

\newcommand{\ud}[1]{\, \mathrm{d}#1}

\renewcommand{\P}{ \mathbb P }

\newcommand{\la}{\langle}
\newcommand{\ra}{\rangle}

\newcommand{\eps}{\epsilon}

\renewcommand{\geq}{\geqslant}

\newcommand{\bs}{\boldsymbol}
\newcommand{\Ub}{\boldsymbol U}
\newcommand{\ub}{\boldsymbol u}
\newcommand{\qb}{\boldsymbol q}

\newcommand{\Zb}{\boldsymbol Z}

\shorttitle{Transport due to Transient Waves}
\shortauthor{J.M. Restrepo, J.M. Ram{\'i}rez}

\title{Transport due to Transient Progressive Waves}

\author{Juan M. Restrepo\aff{1,2} \corresp{\email{restrepo@math.oregonstate.edu}},
  Jorge M. Ram{\'i}rez \aff{3}   
}

\affiliation{
\aff{1}Department of Mathematics, Oregon State University, Corvallis OR 97330 USA
\aff{2} Kavli Institute of Theoretical Physics, University of California at Santa Barbara, Santa Barbara CA 93106 USA.
\aff{3} Departamento de Matem{\'a}ticas, Universidad Nacional de Colombia Sede Medell{\'i}n, Medell{\'i}n Colombia}

\begin{document}

\maketitle

\begin{abstract}
We describe and analyze  the mean  transport due to numerically-generated transient progressive waves, including breaking waves. The waves are packets and are generated   with a boundary-forced air-water two-phase  Navier Stokes solver.
The analysis is done in the Lagrangian frame. The primary aim of this study is to explain how, and in what sense,   the transport generated by transient waves is larger than the transport generated by steady waves.  Focusing on a Lagrangian framework kinematic description of the parcel paths it is clear that the mean transport is well approximated  by an irrotational approximation of the velocity. For large amplitude waves the parcel paths in the neighborhood of the free surface exhibit increased dispersion and lingering transport due to the generation of vorticity.
 Armed with this understanding it is possible to formulate a simple  Lagrangian model which captures the transport qualitatively for a large range of wave amplitudes. The effect of wave breaking on the mean transport is accounted for by parametrizing dispersion via a simple stochastic model
 of the parcel path. The stochastic model is too simple to capture dispersion, however, it offers a good starting point for a more comprehensive   model for mean transport and dispersion.

\end{abstract}


\section{Introduction}
\label{Sec_Intro}

The mean transport due to progressive waves here refers to  the mean Lagrangian velocity generated by the waves. 
 \cite{LH53}  presented a  derivation of the mean Lagrangian transport, due to monochromatic waves,  as an asymptotic series of averaged parcel paths
 and related the terms in the Lagrangian series to a series in the Eulerian framework. The asymptotic parameter was identified as the 
 small distance traveled by a fluid parcel over a period of the wave motion. The first non-zero term in the series of the mean Lagrangian velocity
 is referred to as the {\it Stokes drift velocity}.   (\cite{stokes1847} had derived the lowest-order approximation of the mean transport due to monochromatic progressive waves 
 some nearly a century earlier).  Reviews and analysis of wave-generated residual flows are found in \cite{bremer16} and \cite{bremer17}.
  An alternative derivation of the mean transport can be accomplished via the 
  generalized Lagrangian mean approach \citep{GLM},  wherein the asymptotic series is developed based upon the more geometric displacement vector (see  \cite{OB}, for a detailed description of this approach).

How mean transport due to waves affect the dynamics of oceans at time and space scales larger than waves has been the subject of considerable attention.   Longuet-Higgins and collaborators ({\it cf}, \cite{LHS60}, \cite{LHS62}, \cite{LHS64}, \cite{L70})  proposed a radiation stress to describe wave generated transport and examined ways in which  the stress or the interaction of the stress with a background current plays out in a variety of different important geophysical flows.
The interaction of the residual flow due to gravity waves and the mean flow  that makes up what is known as Craik-Leibovich (CL) theory (see \cite{L83} for a review)
provides an alternative {\it vortex force} formulation to the radiation stress for the coupling. The vortex force enters the momentum equation as a contribution to the  cross product of the 
total fluid rotation with the transport velocity, as approximated by the mean Eulerian velocity plus the Stokes Drift velocity. A Bernoulli term also enters the momentum equation as an adjustment to the pressure gradient.  (Similarities and differences between the radiation stress formulation and the vortex force
formulation are detailed in \cite{LRM06}).  The destabilizing effect of the waves on currents in the CL theory 
 forms the conceptual basis for the generation of  Langmuir circulations,  later generalized in \cite{MSM97} to describe Langmuir turbulence.
 The CL theory was extended to capture the  wave-driven circulation, in  \cite{mr99} and forms the basis for a shallow-water conservative dynamic of waves and currents \citep{mrl04}.

  This  work  is concerned with   the mean transport
due to transient progressive waves of small as well as of  large amplitude. To this end, we will be applying ensemble characterizations to describe the average transport. We will also make use of the insights gained from the analysis of parcel paths to propose a simple model for the averaged transport.

The analysis will focus on characterizing and interpreting the transport
of  a specific set of numerically-generated experiments of progressive waves, including breaking waves. 
The numerically-generated transport considered in this paper revisits the transport results reported in 
 \cite{deikepizzomelville}, hereon, DPM17. We use the same data, in fact.
 The numerical data  are solutions to the boundary-forced Navier-Stokes equations, approximated by numerical means,  for a heavy fluid (water) and an overlying light fluid (air). 
 The waves were generated by a transient forcing boundary condition and were absorbed by the opposing boundary condition. Hence, we do
not encounter some of the difficulties that arise in the real setting with regard to defining a mean transport, where waves may not have clean starting or ending times.
  We will  make use of a Lagrangian description of the fluid flow in order to characterize and understand the transport.  
 The Lagrangian parcel paths are computed using interpolation from the grid Eulerian velocity. 
 The focus in DPM17 is on the transport due to wave breaking progressive waves and in the phenomenon of wave riding, in particular. We will touch upon this topic  but our emphasis will be on characterizing transport  of small and large transient progressive waves, as a function of depth and of wave slope, a parameter that controls the wave amplitude generated by the time dependent boundary condition. 
 
  The transient progressive waves we consider are focusing wave packets. Mean transport due to wave packets is derived in \cite{bremer17}. However, the data considered here   are not  amenable to the analysis presented in \cite{bremer16} concerning wave groups,  because they do not have the requisite time scale separation. We will demonstrate that the estimate that leads to the classical Stokes drift formula, which  relies critically on an assumption in the wave statistics,  does not hold generally  for the numerically-generated transient waves under consideration. This is at the heart of the explanation of  why transient waves may produce significantly more transport than progressive waves.

 The numerical data and its generation appear in Section \ref{sec:num}. In Section \ref{sec:background} transport due to progressive monochromatic waves is summarized. Doing so gives us the opportunity to focus on the key distinction between transient and steady progressive wave transport,  namely, an assumption critical for approximating the transient transport in terms of the series expansion leading to the familiar progressive wave transport derived in \cite{LH53}, as modified in \cite{r07} to include unresolved processes parametrized by diffusion processes. Section \ref{sec:kinematics} 
 describes how transient mean transport is computed from the numerical data and proceeds to describe the transient transport. We contrast our analysis from the one provided in DPM17, which is based upon the same data. The kinematics of the parcels suggests a 2-parameter model for the parcel dynamics. The model is described in Section \ref{sec:model} and when tuned, recovers the mean transport and the dispersion in the data. The model is based upon the parcel  kinematics and thus incorporates the fundamental aspects of the parcel dynamic that contribute to the transport. A summary of results appears in Section \ref{sec:summary}.

 \section{Generation of Numerical Progressive Wave  Lagrangian Paths}
\label{sec:num}

In DPM17  the authors employ Gerris  (see, \cite{popinet09}), a Navier-Stokes equation solver, to obtain approximations of the motion of an  air/water fluid under the action of a downward gravity force with magnitude $g$.  Time is denoted by $t \ge0 $ and the simulation runs until $t=T_f=35$ s. The computations are done in two space dimensions with transverse coordinate denoted by $x$;  $z$ is the vertical coordinate, which increases upward from the quiescent reference level, $z=0$. The `tank' extent is 24m, and the depth of the water-filled tank is $h=1$m. The air/water interface has surface tension. The fluid is  subjected to a time dependent `paddle'  forcing boundary condition at  $x=0$ generating a wave  packet which dissipates at the other end of the tank by absorbing boundary conditions.  Zero velocity boundary conditions are imposed at the bottom of the tank, $z=-h$, and at $z=h$, the top of the domain. The Navier-Stokes solver uses the fresh water and air kinematic viscosity ratio and the computations reach Reynolds numbers in the order of 40,000 (further details of the numerical generation of the flow  are found in DPM17). 
At rest, initial conditions are invoked in all of the numerical simulations. The Eulerian velocity is denoted by ${\bf q}(x,z,t)$,  and the free surface is $z=\eta(x,t)$.

Throughout this study we will make reference to the following, which we denote as the `linearized wave solution':
\begin{eqnarray}
\eta^w(x,t) &=& \sum_{n=1}^{N} a_n \cos(k_n (x-x_b) - \omega_n t)), \nonumber \\
\ub^w(x,z,t) &=&  \nabla \phi(x,z,t), \quad \mbox{where} \nonumber \\
\phi(x,z,t) &=& - \sum_{n=1}^{N} \frac{a_n \omega_n}{k_n} \frac{\cosh(k_n(z+h))}{\sinh(k_nh)}
\sin(k_n (x-x_b) - \omega_n t),
\label{paddle}
\end{eqnarray}
where $x_b=12$m, $t_b=25$s are respectively, the `focusing' position and time.  The velocity $\ub^w$ is an Eulerian velocity, and $\phi$ and $\eta^w$ are the velocity potential and sea elevation that describe  an irrotational, incompressible infinitesimal-amplitude progressive wave packet.

In the simulations, the paddle is driven by the vector field $\ub^w(0,z,t)$ in (\ref{paddle}) with $N=32$. The simulation used the dispersion relation for  angular frequencies $\omega_n = \sqrt{g k_n \tanh( k_n h)}$,  where the  $k_n$  are the wavenumbers.  The paddle amplitudes $a_n$  were prescribed as follows:
DPM17 employ the  {\it slope} $S$, as an ordering   parameter in the simulations.  $S 
=\sum_{n=1}^N k_n a_n = N s$, and $s = k_n a_n$, constant, $n=0,1,...,N-1$.
In the specification of each run the slope $S$ is fixed. The data includes simulations for the following cases
\[ S \in \{0.16, 0.192, 0.256, 0.288, 0.32, 0.336, 0.352, 0.368, 0.384, 0.4, 0.416\}, \]
with breaking occurring for $ S \geq S_0 = 0.336 $. The wavenumbers $k_n$ were found using the dispersion relation and the frequencies chosen as follows:  The component frequencies $\nu_n=0.5458+n \delta_\nu$ Hz, where $\delta_\nu = 0.0222$  Hz.
 The central frequency is denoted $\nu_c = \omega_c/2 \pi = 0.89$ Hz.
  
\begin{figure}\centering
	\includegraphics[width=5.1in,height=1.5in]{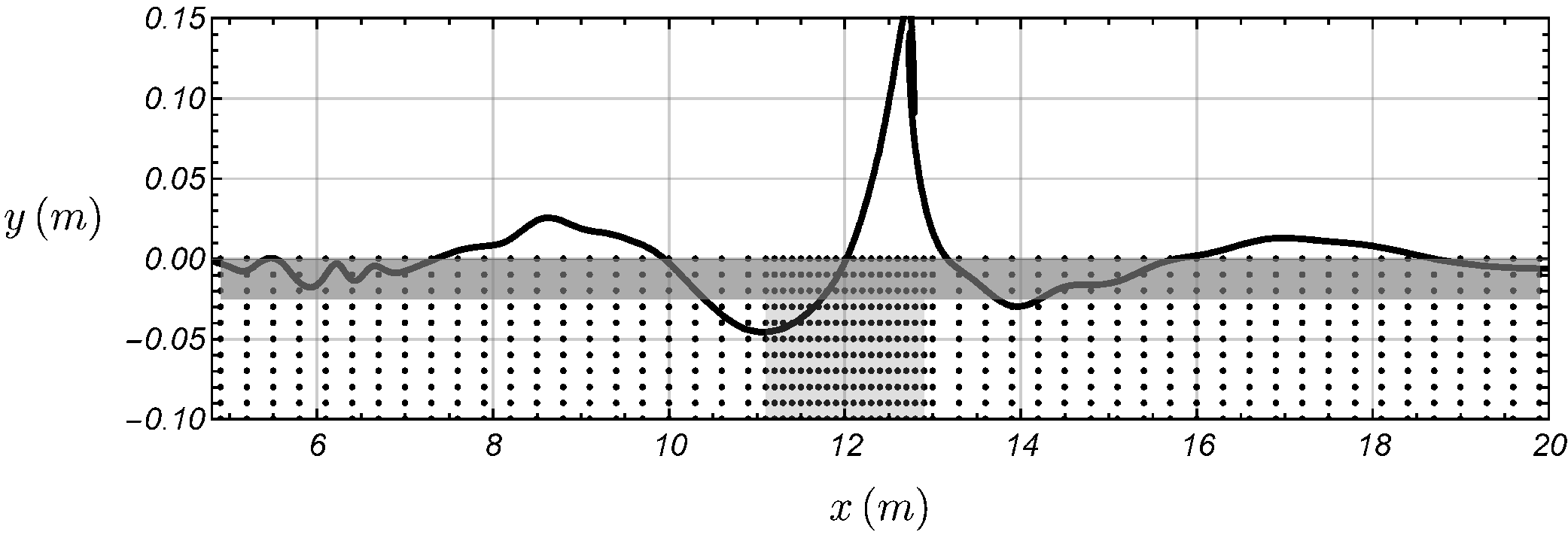}
	\caption{Detail of the numerical tank, featuring the  initial positions $\Zb(0)$ of  all of the parcel paths (dots) used in the analyses. The free surface of the wave right before breaking is shown, for the case $ S=0.38 $.  The region with parcels with initial positions belonging to the near-surface (approximately no deeper than 3cm below the quiescent surface) is shaded in   gray. Parcels starting in the upper gray zone
	may spend some portion of their history on the sea surface, or for some large values of $S$, ejected during a breaking wave event.
	 The zone that marks the boundary of parcel paths staring in the zone  $ |X(0) - x_b|<\frac{\lambda}{2} $ is shown in lighter gray, has a higher density of parcel paths. }
	 \label{Fig_PointsTank}
\end{figure}

Parcel paths $\Zb(t)=(X(t),Z(t))$ were computed by an explicit first order in time tracer advection scheme using the velocity field from the Navier-Stokes solver, using second-order interpolation. The numerics approximately solve the parcel path equation 
\begin{equation}
\dot{\Zb}(t) = {\bf q}(\Zb(t),t).
\label{transport}
\end{equation}
Figure \ref{Fig_PointsTank} shows the starting locations of all paths $\Zb(t=0)$  considered in the analyses that follow. 
 Throughout this work we will connote the numerically-generated outcomes as {\it data}. 
The shading scheme organizes the parcels by their position at the  starting time. Our analyses focus on paths with $Z(0)$ within 10 cm of the surface. The gray area nearest to the surface will be henceforth referred to as the `near-surface' paths. For large $S$ some parcel paths within the near-surface, exhibit complex  dynamics such as  ejection and wave riding, the latter of these dynamics  analyzed in \cite{pizzo17}. For the diagnosis of transport leading to Figure \ref{Fig_DeltaXvsUst}, all of the points shown in Figure  \ref{Fig_PointsTank} were taken into account. In formulating a model for transport and dispersion (see Section \ref{sec:model}), we focused on paths starting within the gray area in the middle, $|X(0) - x_b|<\lambda/2$ where $\lambda = 1.96$ cm is the wave period.

\section{Transport Due to Breaking Monochromatic Progressive Waves}
\label{sec:background}

In order to contrast the transient case we first review transport dynamics under the action of steady monochromatic progressive waves. This review
 will also prove helpful in reinterpreting the results on transport due to wave breaking presented in DPM17.
 
 The Lagrangian  velocity will be denoted by $ \Ub(t) = \dot{\Zb} =(U(t),W(t)) $.
The Stokes drift velocity is the lowest-order estimate of the mean Lagrangian velocity  due to time periodic irrotational infinitesimal waves.  For progressive waves of amplitude $a_c$, the Stokes drift is constant and proportional to $a_c^2$ (see \cite{LH53}). 
In \cite{r07} this result was extended to include an additive   stochastic velocity component. 
When properly non-dimensionalized, the Eulerian velocity  is decomposed as 
\[
\qb(x,z,t) = \qb^{D}(x,z,t) + \eps \ub^w(x,z,t) + \eps^2 \qb^C(x,z,t,X,T),
\]
where $\qb^{D}$ is a zero-mean velocity associated with unresolved stochastic 
sub-wave velocity contributions, including contributions due to wave breaking,   $ \ub^w $ is the irrotational or wave component of the velocity, whilst $ \qb^C $ represents the rotational component (possibly including an imposed  current ${\bf v}^C$)  as well as the  
velocity associated with transport due to waves and the residual flow due to breaking waves. This decomposition is based upon the assumption  that wave orbital velocities are larger than currents
(see \cite{mr99} and \citealt{RMMB}), and diffusive scale velocities, larger than wave orbital velocities. The stochastic component is a parametrization of unresolved processes.  We will adopt a Wiener process to model this velocity (clearly, an ad-hoc representation that can elicit a number of objections, the least of which is that it is incompressible only in the mean).
 Associated with the above velocity decomposition, the parcel path decomposes as 
 $
	\Zb = \Zb_0 + \eps \Zb_1 + \eps^2 \Zb_2 + \cdots, 
	$
 the various terms  $\Zb_i$, $i=0,1,...$, are assumed to be scaled dimensionally to unity.

We use the operator $\langle \cdot \rangle$ to denote \emph{ensemble average}, which in this case means expectation with respect to the stochastic component of $\qb$, a Wiener noise process $W(t)$ satisfying $\langle W(t) \rangle =0$, and  $\langle W(t) W(t') \rangle =t \, D  \delta_{t,t'}$ and affecting $\qb$ at the fastest time scales. Further, for some $f(t)$,  $ \overline{f}$ denotes the average of $f$ over the period $T_c=1/\nu_c$. 

The critical observation in the \cite{LH53} analysis is that, in the case of a monochromatic progressive wave, say  $\ub^w(x,z,t)=-\nabla   \frac{a_c \omega_c}{k_c} \frac{\cosh(k_c(z+h))}{\cosh(k_ch)}
\cos(k_c (x-x_b) - \omega_c t)$, the  average  $\overline{\ub^w}$ vanishes. The lowest order Lagrangian and Eulerian velocities also vanish \textbf{in average}. Hence
\begin{eqnarray}
 \left \langle \overline{d\Zb_0  } \right \rangle&=& \left \langle \overline{ \sqrt{2D}  d {\bf w}_t } \right \rangle =0, \label{z0dot} \\
 \left \langle \overline{d\Zb_1 } \right \rangle &=&  \left \langle \overline{ \ub^{w}(\Zb_0,t) \ud t} \right \rangle =0, \label{z1dot}\\
 \left \langle \overline{d\Zb_2  } \right \rangle & =& \bs{v}^C dt + \ub^{St} \ud t,
\label{Z2}
\end{eqnarray}
where $d {\bf w}_t :=(d W_t, 0)$.
The third equation in (\ref{Z2}) has the imposed current (if present) and a residual flow due to wave breaking,   as well as the Stokes drift velocity $\ub^{St}$. The Stokes drift $\ub^{St}=(u_c^{St},0)$ can be obtained  by computing 
\[
u_c^{St} = \left \langle \overline{ \int_0^t  \ub^w(\Zb_0,s) \, ds \cdot \nabla \ub^w(\Zb_0,t)} \right \rangle,
 \]
 where $\Zb_0 = \Zb_0(0) + \sqrt{2D} \, {\bf w}_t $. For progressive monochromatic shallow-water waves, the steady  Stokes drift velocity is  
\begin{equation}
	u_c^{St} =    \frac{S_c^2 c_p}{2 }  {\cal D} \, \cosh^2 [  k_c (z+h)] \mbox{csch}(2 k_c h),
	\label{Def_Ust}
\end{equation}
 with $S_c= k_c a_c$, $c_p=\omega_c/k_c$ is the phase speed, and
${\cal D} = \frac{1}{1+\Delta}$, where $\Delta = \frac{k_c^4 D^2}{\omega_c^2}$.      The Stokes drift velocity 
is proportional to  $a_c^2$ via $S_c^2$. Note that the stochastic term leads to an increase in wave dispersion. It also leads to 
a suppression of wave-generated transport (see \cite{r07}).
 ${\cal D}$ is 1 when the stochastic process is zero, leading to the familiar expression for the Stokes drift velocity under progressive monochromatic waves, obeying the shallow-water dispersion relation. 
Condition (\ref{z1dot}) holds exactly for monochromatic waves and   holds as well  for random-phase, stationary linear waves (see \cite{Huang70}). 
%It does not hold for the transient from which our data was collected. 

\section{Transport due to Transient  Waves}
\label{sec:kinematics}

In the analysis of the data,  the {\it mean horizontal displacement} is defined as $\langle X(T_f)-X(0) \rangle$. We also define {\it transport} as the mean horizontal velocity $\frac{\langle X(T_f)-X(0) \rangle}{T_f}$ over the whole simulation.
In this context, the ensemble average $\la \cdot \ra$ of a quantity is estimated as the average of such quantity over all paths with initial location within the  central light shaded rectangle of black points in Figure \ref{Fig_PointsTank}.   The assumption here is that the uncertainty modeled as a stochastic component in $\bf q$, translates into the data as uncertainty over the parcel paths initial condition $\Zb(0)$.

In what follows we describe  essential aspects of the  kinematics of the  transport. We will use dimensional quantities.
The average transport is  obtained by averaging (\ref{transport}). 
For transient waves, meaning for waves for which (\ref{z1dot}) does not hold over the interval of time of interest, (\ref{z0dot})-(\ref{Z2}) 
does not apply and the Stokes drift, as defined above,  makes no sense. 

We found that by setting  ${\bf q} \approx \ub^w$, as given in (\ref{paddle}), into (\ref{transport}) we were able to construct parcel paths that were qualitatively very similar to the data, regardless of which initial parcel path position we chose. The paths so constructed mimic the details in space and in time, for a large range of $S$, excluding breaking cases. The conclusion to be drawn from this is that, if dissipation and dispersion are appropriately parametrized, a linearized velocity model can capture most of the details of the parcel paths, in spite of the fact that the numerical solutions are those approximating the Navier-Stokes equations. 

The fact that ${\bf q} \approx \ub^w$ approximates the Lagrangian velocity implies that the horizontal displacement $ dX $ is proportional to $S dt$, regardless of the value of $S$. As DPM17 found, when $S$ is small, we found that  the \textit{mean transport} is proportional to $S^2$, and when $S$ is large, the \textit{mean transport} can be significantly larger, proportional to $S$. This is explained as follows:  when $S$ is small, (\ref{z1dot}) is approximately satisfied. For example, near to the sea surface and for $S=0.16$, $\overline{U(t)} \approx 10^{-6}$ m/s. Hence  (\ref{z0dot})-(\ref{Z2}) approximately holds and the Stokes drift is an approximation of the mean transport which is thus  approximately proportional to $S^2$ as per equation \ref{Def_Ust}. However, for $S=0.32$, $\overline{U(t)} \approx 10^{-3}$ m/s. In this case  the mean transport is proportional to the  amplitude of $\ub^w$ ({\it i.e.}, to $S$) and is given by the average of $ \dot{\Zb}(t) $ in (\ref{transport}).  This $S$ dependency holds for mean transport at any depth.

\begin{figure}
 \centering
(a) \includegraphics[width=2.4in,height=1.7in]{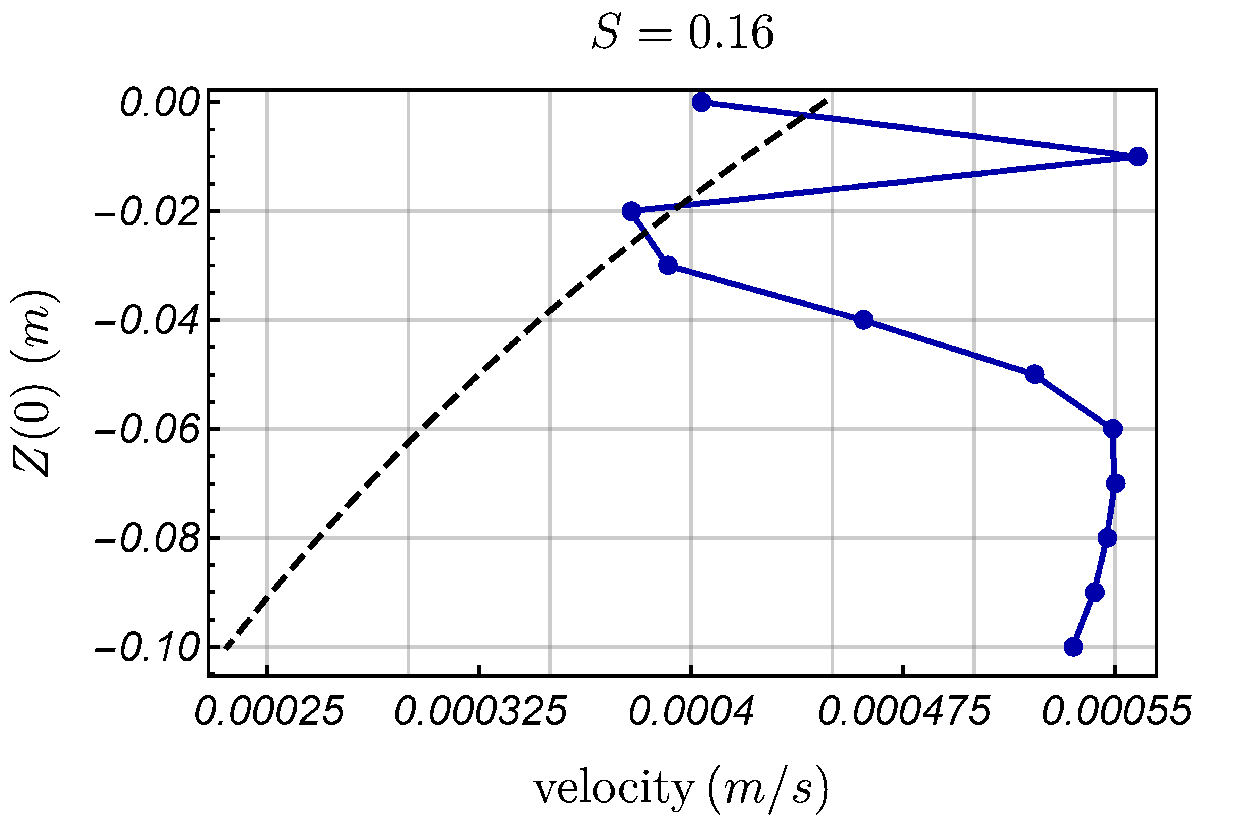}
 (b)\includegraphics[width=2.4in,height=1.7in]{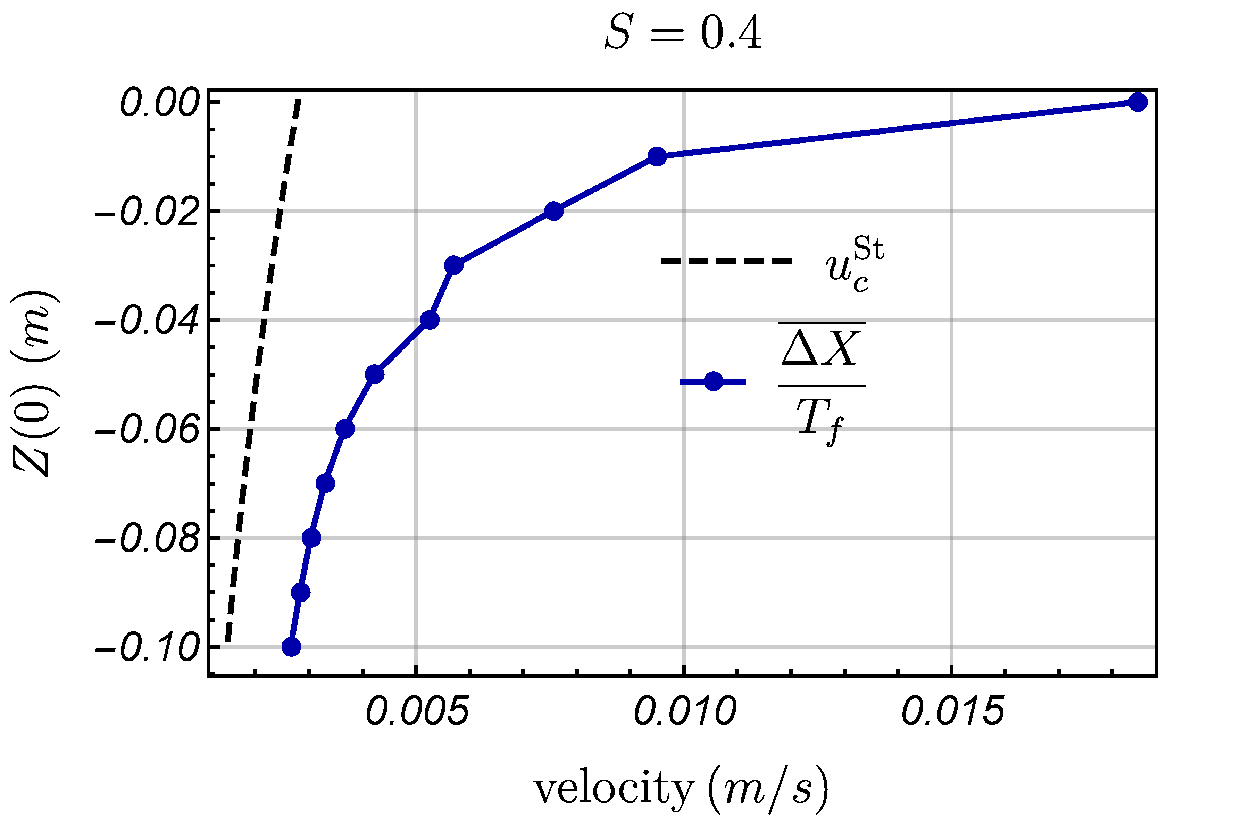}
	\caption{ Mean horizontal transport as a function of depth, compared with the Stokes drift (dashed)  for the cases (a) $S=0.16$ and (b) $S=0.4$, a breaking case. $T_f = 35$s  and the mean here is taken over all paths with same initial depth. The Stokes drift velocity $u_c^{St}$ is computed using (\ref{Def_Ust}) with $ {\cal D} = 1 $. (Note horizontal scales). The small $S$ case is coincidentally well estimated by
	the Stokes drift. This is not true for the large $S$ case, breaking or otherwise.}  
		\label{Fig_DeltaXvsUst}
\end{figure}

Figure \ref{Fig_DeltaXvsUst} depicts the mean transport  and the Stokes drift velocity  as a function of depth for two values of $S$.  Note that for large $S$ the transport can be three times larger in the near-surface region when compared with deeper paths. For  small $S$,  $\overline{U(t)} \approx 0$ which is consistent with the parcel paths being very similar to  the familiar-looking paths of monochromatic waves, and thus the mean transport can indeed be approximated by $u_c^{St}$. 

\begin{figure}
\centering
		\includegraphics[width=5.3in,height=2.8in]{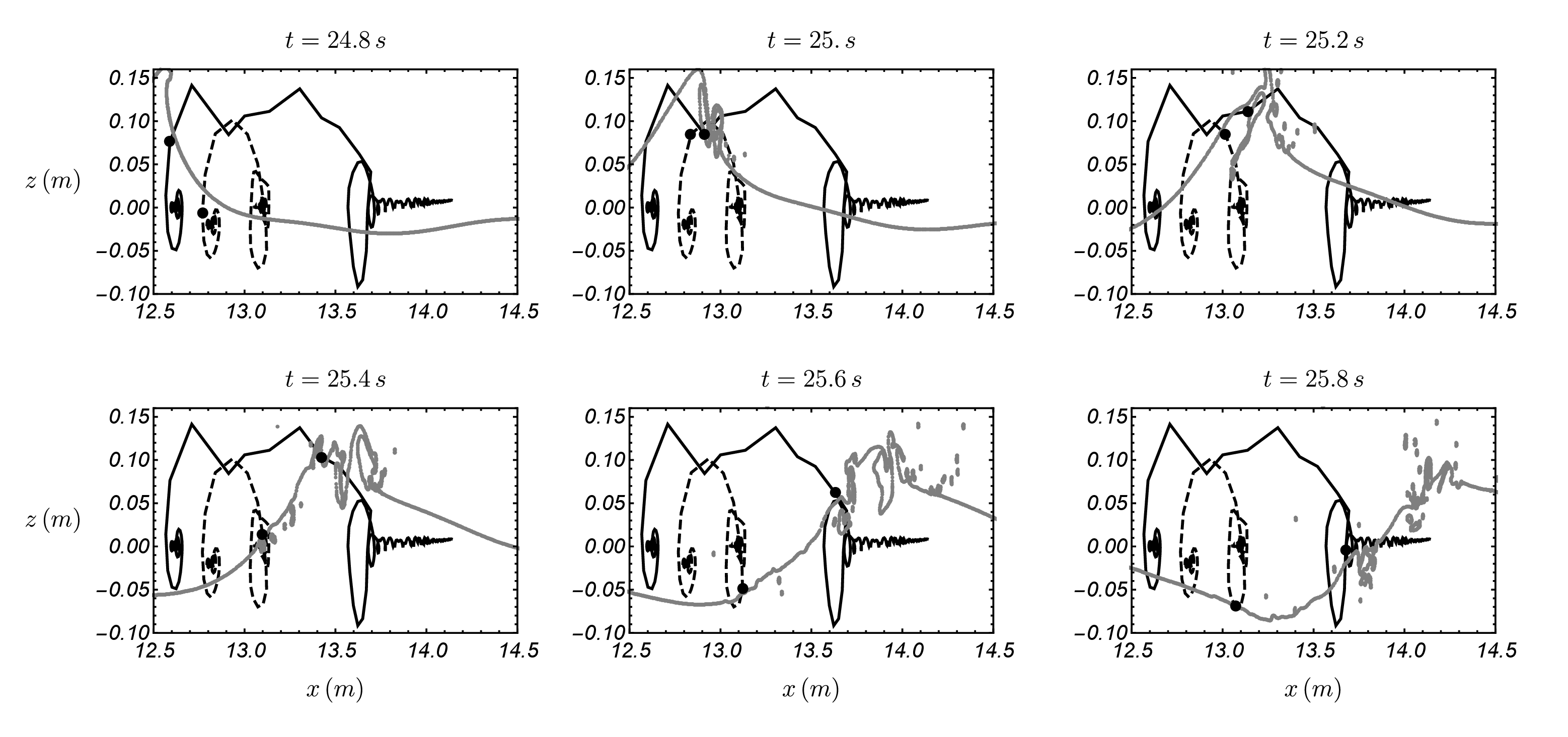}
	\caption{Lagrangian parcel path histories.   $S=0.4$, The breaking free surface is shown in gray.  The smoother path (dashed line) that follows an elliptical path punctuated by a large displacement is most typical of  parcel paths, thoughout.  The more complex (solid line) path is of the type we denote as  a {\it rider} parcel path, since it spends some time on the sea surface. It started near $x_b$ on the sea surface.}
	\label{Fig_PathTypes}
\end{figure}

\begin{figure}
\centering
		\includegraphics[width=5.4in,height=2.7in]{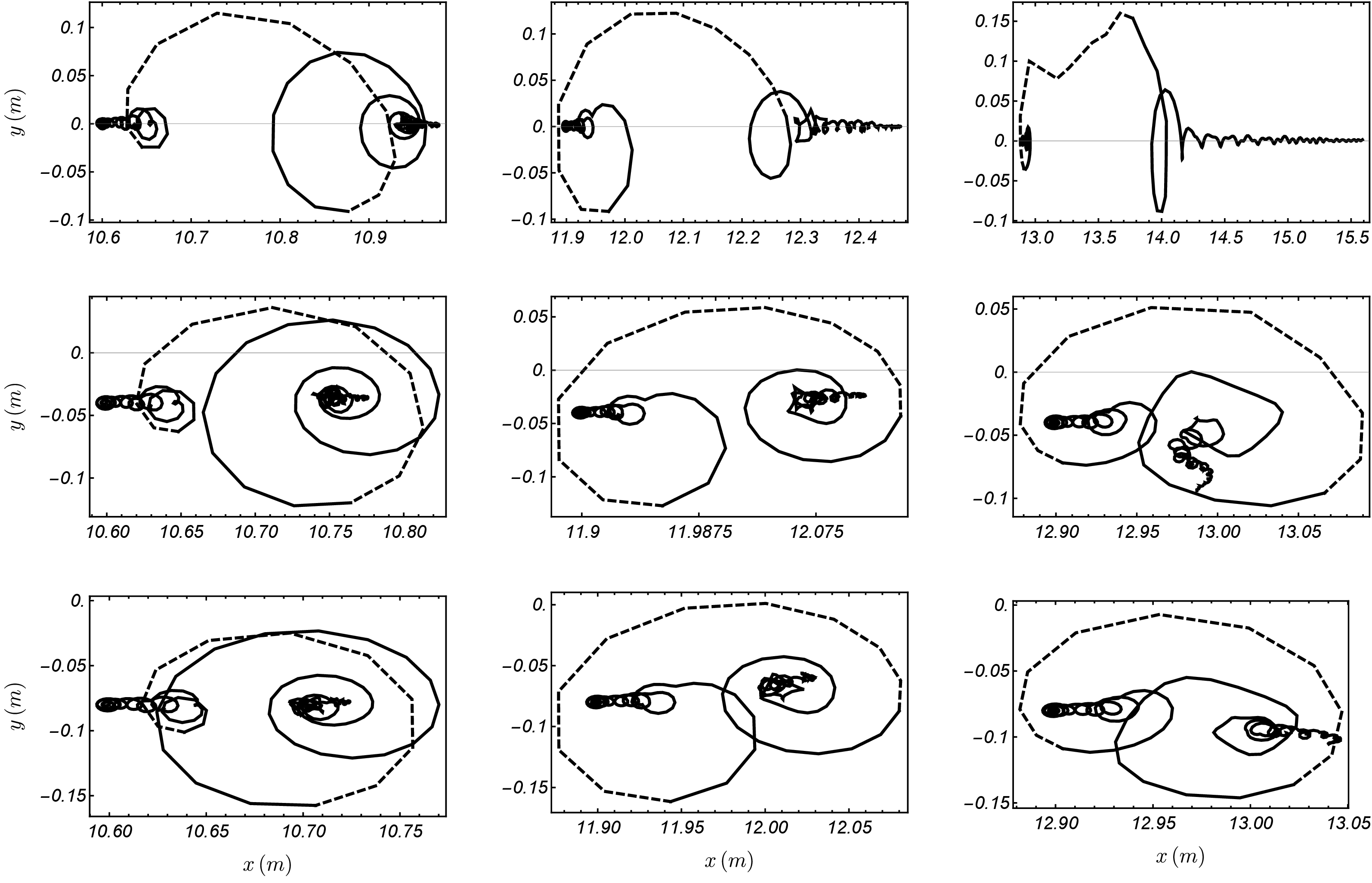}
	\caption{Lagrangian parcel path histories corresponding to  different initial locations. The wave slope $S=0.4$. 
	Highlighted as dashed is the portion of the path  when the crest of the wave passes overhead. } 
	\label{Fig_ElongatedEllipses}
\end{figure}

\begin{figure}\centering
	\includegraphics[width=5.6in,height=3in]{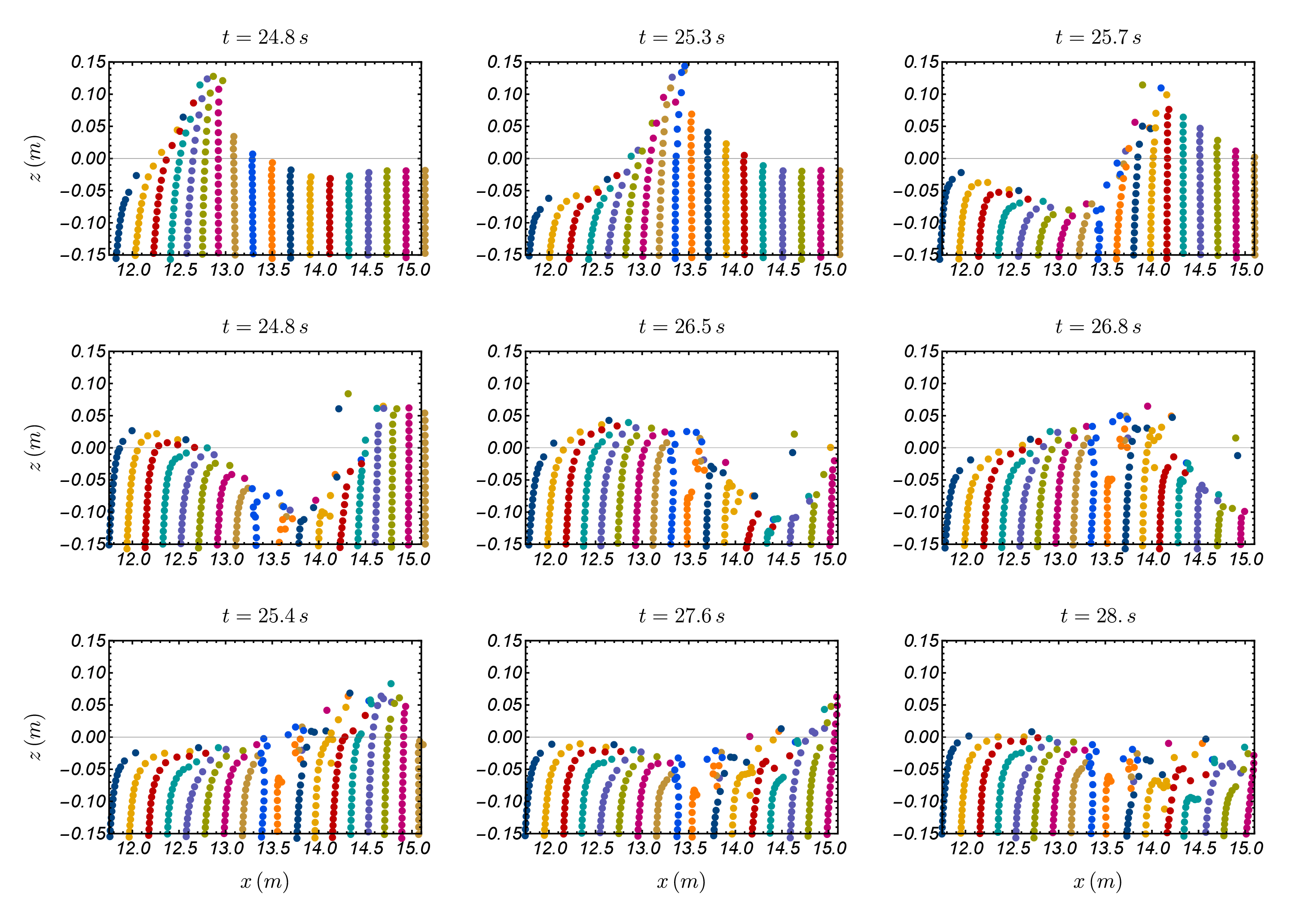}
	\caption{Evolution of parcel  paths. $S=0.4$. Transport and dispersion is  evidenced by irreversible forward displacement of the  parcels. Complex-motion  displacements are found to be confined to a very thin layer close to the free surface.} 
	\label{Fig_SummaryTransport}
\end{figure}

Figure \ref{Fig_PathTypes} shows a couple of parcel paths for large $S$, {\it i.e.}, highly transient waves. (See also Figure \ref{Fig_SummaryTransport} and Figure \ref{Fig_ElongatedEllipses}).
 The gray line demarcates the (breaking) free surface. A dot indicates the parcel position at the time specified by the frame. 
 The solid path corresponds to a wave-riding parcel. These parcels spend some  time on the free surface, resulting in a complex displacement history (see \cite{pizzo17}). The parcel path marked as a dashed line is taken from an initial condition with $Z(0) = -0.04$. The large excursion coincides with the passage of the wave. The dashed path is more typical of the parcel path dynamics inside the water column for large $S$ including breaking waves. 
 Figure \ref{Fig_ElongatedEllipses} features several parcel paths for the $S=0.4$ (breaking) case in different regions within the tank. The first row corresponds to parcel paths starting at the surface. Paths in the two lower rows correspond to parcel paths originating below the \textbf{near-surface} region in Figure \ref{Fig_PointsTank} and are very typical of the paths under large unbreaking  as well as breaking waves. The dashed portion  highlights the  parcel flight that coincides with the wave passing overhead. Clearly, it is this large amplitude displacement and the residual motion after the wave passage  that largely accounts for significant mean transport.As a function of depth $z$ the dashed  paths scale proportional to the average of $\exp(k Z(t))$, where $Z(t)$ is the vertical component of a specific paths displacement over time. This exponential drop with depth is inherent in  (\ref{paddle}).

If $S>S_0$, wave breaking can occur and vorticity is generated in the flow but found  confined to a very thin layer close to the sea surface. This vorticity lingers after the wave passes. Wave breaking  enhances dispersion, regardless of the manner in which the waves break,  but it is the ensuing vorticity that  enhances transport: wave momentum is transferred to the mean flow.  See Figure \ref{Fig_SummaryTransport}.

\begin{figure}
	\centering
	\includegraphics[width=5in,height=1.5in]{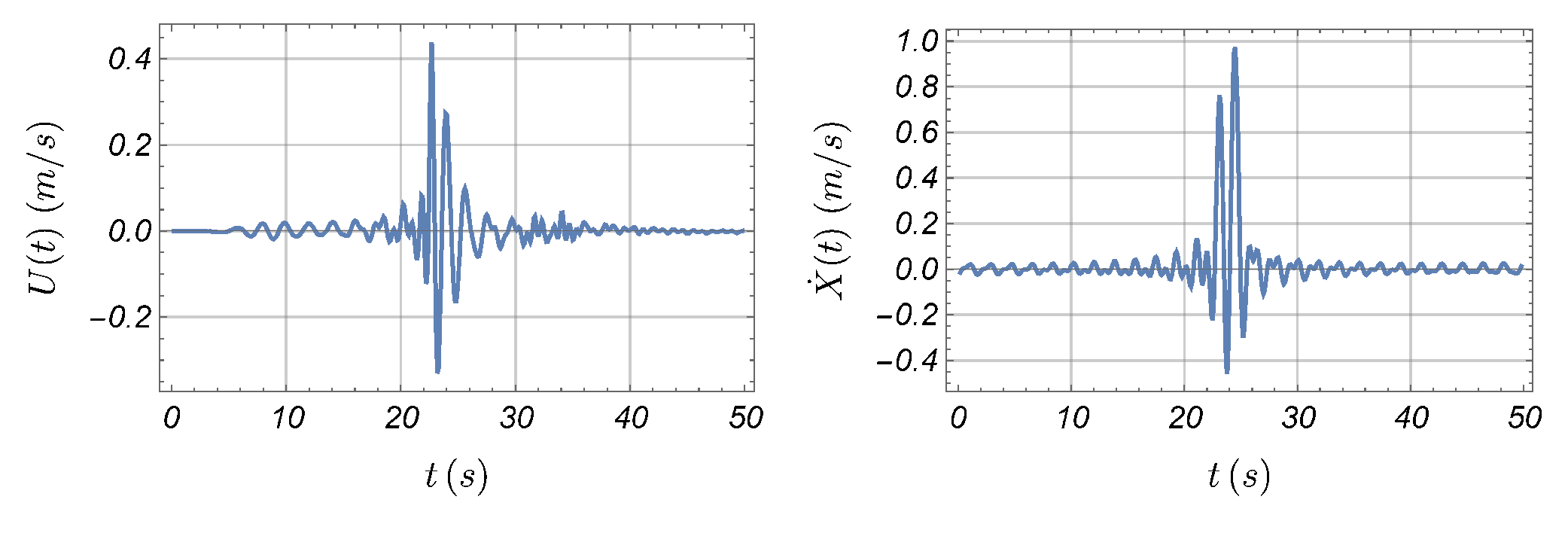}
	\caption{ Left panel: horizontal Lagrangian velocity (data)  of a  parcel with $ \Zb(0)=(10.6,-0.07) $ in the case $ S=0.4 $. Right panel: horizontal Lagrangian velocity, obtained by numerically solving $ \dot{\Zb} = \ub^w(\Zb(t))  $ with $ \ub^w $ given by (\ref{paddle}). Note the difference in vertical scale. The spectrum of the forcing boundary condition creates a transient velocity everywhere in the tank.}
	\label{transientvelo}
\end{figure}

A time series of the horizontal component of the velocity, following one of the parcels is shown in the left panel of Figure \ref{transientvelo}.  The  right panel of Figure \ref{transientvelo} depicts  the horizontal velocity obtained from  solving 
\begin{equation}
 \dot{\Zb} = \ub^w(\Zb(t)), 
 \label{doteq}
 \end{equation}
  with $ \ub^w $ given by the linearized wave field in (\ref{paddle}), and $ Z(0)=-0.07 $ m. The two are qualitatively similar, however, the time series associated with (\ref{doteq}) severely overestimates the data velocity for large $S$.
  Evidently, surface breaking, and the generation of vorticity/dissipation inherent in the Navier-Stokes solution has a significant  effect on the amplitude of the parcel path. Equation (\ref{doteq})  forms the basis for a transient transport model to be introduced in a subsequent Section. 
A simple parametric model for dissipation is proposed to improve the compatibility of (\ref{doteq}) with regard to predicting mean transport at
any depth for any $S$. We propose a modification to (\ref{paddle}) that includes dissipation parametrically as follows 
\begin{eqnarray}
 \dot{\Zb} &=& \ub_d^w(\Zb(t)) =  \nabla \phi_d(\Zb(t),t),  \label{doteqdiss} \\
 \phi_d &=& - \sum_{n=1}^{N} \frac{a_n \omega_n}{k_n} \frac{\cosh(k_n(z+h))}{\sinh(k_nh)}
 \sin(k_n (x-x_b) - \omega_n t) e^{-\beta  k_n^2 t},
 \label{paddled}
 \end{eqnarray}
where $\beta$ is a tunable dissipation parameter. In Section \ref{sec:model} we provide further details on how this parameter is chosen.
  The essential lesson is that even though the numerical code is solving Navier-Stokes, for small and large amplitude waves, a simple model for the Lagrangian velocity captures the paths qualitatively most everywhere in the tank at any time and can be tuned to deliver reasonable 
  mean transport estimates for a large range of $S$.   How close the model captures the mean transport will be touched upon in Section \ref{sec:model}, when inclusion of wave breaking dispersion is taken into account.

\begin{figure}
	\centering
		\includegraphics[width=5in,height=2in]{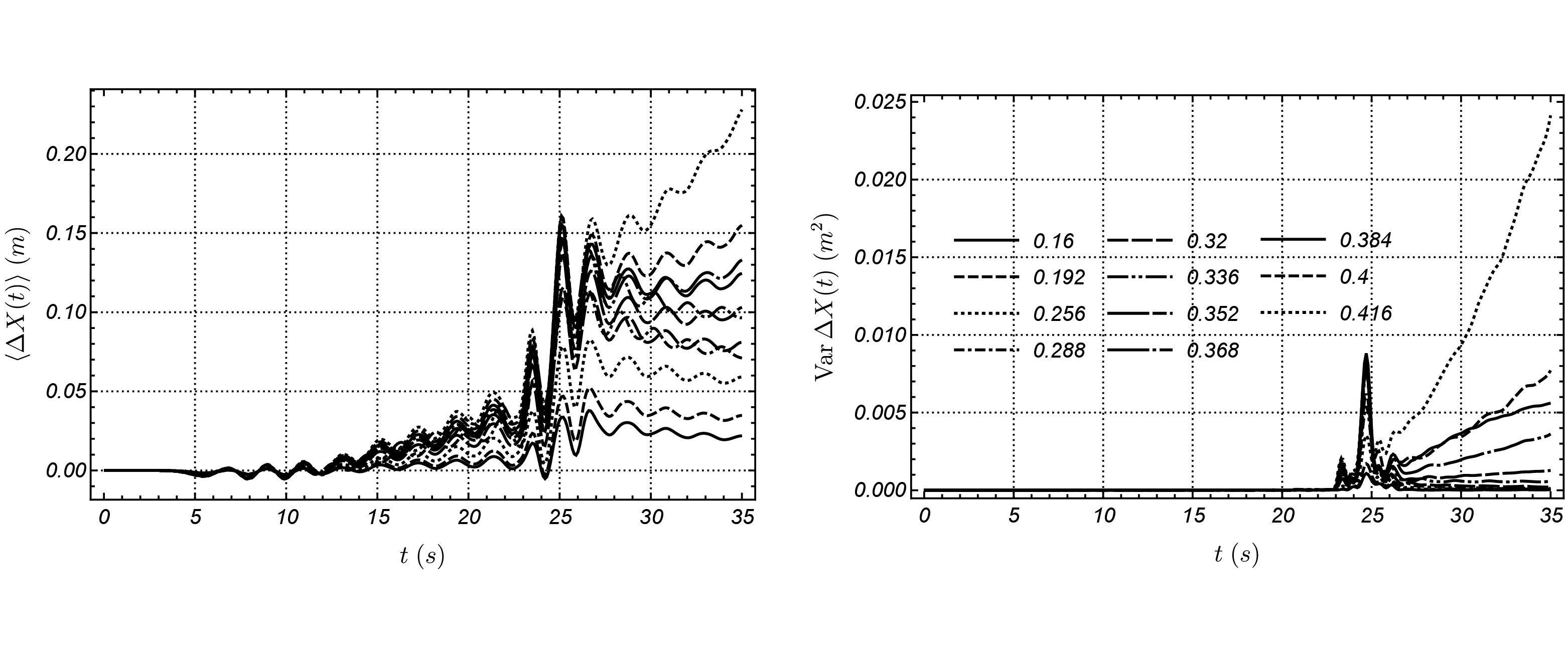}
	\caption{As a function of $S$, the mean (left) and variance (right) of  the horizontal displacement $ \Delta X(t) = X(t) - X(0) $. Averages are taken over the trajectories starting in points within the gray-shaded middle area of Figure \ref{Fig_PointsTank}. The breaking cases correspond to $S>S_0=0.336$. The dispersion persists after the wave breaks, for breaking cases.}
	\label{meanvar}
\end{figure}

Figure \ref{meanvar} shows the time dependent mean and variance of the (horizontal) displacement, estimated from data. Large
horizontal displacements are caused by large amplitude and highly transient waves. On the right we present the horizontal displacement variance. Before the wave passes overhead, the variance is nearly zero. Once the wave passes overhead, there is an injection of variance. For non-breaking cases, the variance reverts back to nearly zero, after the wave passes. However, this is not the case for breaking cases: the variance grows linearly, in fact. Wave breaking generates  variability in the horizontal transport, coinciding with increased vorticity.

\section{The Stochastic Model}
 \label{sec:model}
 
\begin{figure}
	\centering
\includegraphics[width=4.4in,height=1.8in]{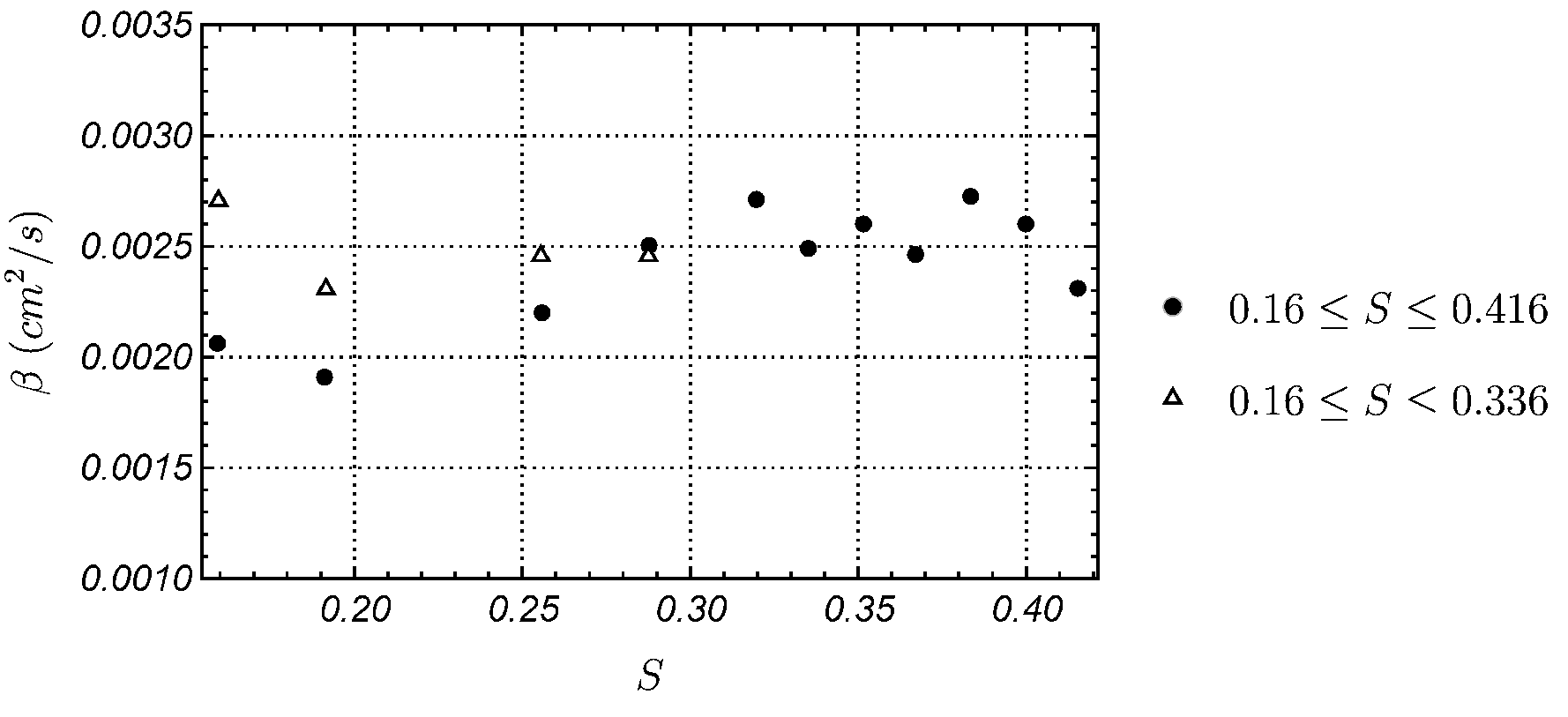}
	\caption{Fitted $\beta$ values as a function of $S$. For non-breaking cases (triangles) we included paths up to the surface in the estimate of $ \beta $. Dots mark the estimates of $ \beta $ for all values of $ S $ and paths initially in the central gray-shaded area of figure \ref{Fig_PointsTank}.}
\label{betafit}
\end{figure}

 The goal in this section is to extend the model (\ref{doteqdiss}-\ref{paddled}) to increase the range of $S$ for which the model can capture  the mean transport due to transient progressive waves, to include breaking waves.  The model 
 is tuned to deliver the correct mean transport but will be based on good qualitative approximations of the  parcel paths, for all values of $S$. 
 That is,  a good approximation of the mean transport for any $S$,  regardless of whether  $\overline{\Zb_1}$ is small
 or otherwise.   A model, consistent  
with the  observations made in Section \ref{sec:kinematics}, is 
 \begin{equation}
\langle \ud\Zb \rangle = {\bf v}^C \ud t +  \langle  \ub_d^w(\Zb,t) \ud t + \sqrt{2 D(t-t_b)} \ud{\bf w}_t \rangle.
\label{ZR}
\end{equation}
The first term on the right hand side is zero in the data, but otherwise,  would represent an imposed current. 
The model has two fitting parameters: 
the dissipation rate $ \beta $  and the molecular diffusion $ D(t) = D_b\,  \Theta(t-t_b) \Theta(S-S_0)$, where $\Theta$ is the Heaviside function, $S_0$ is the threshold $S$ for wave breaking, and $D_b>0$ . The  parameters $\beta$ and $D_b$ were estimated from data:  we used the data to estimate  the ensemble average $ \la X(T_f)- X(0)\ra$ and then  solved
$\ud \tilde{\Zb} = \ub_d^w(\tilde{\Zb},t) \ud t$ for the same initial locations, and numerically estimated the value of $ \beta $ such that $ \la \tilde{X}(T_f)- \tilde{X}(0)\ra$ gave the best approximation to the average from the data. 
Figure \ref{betafit} compares the fit of  $\beta$, as a function of $S$, when the model is tuned to data that includes near-surface 
paths (triangles) and when it does not (circles). The values are similar.  The fit leads to $\beta \approx 0.0024$ m$^2$/s.
 The value of $ D_b(S) $ was estimated following the same fitting methodology using instead the second moment  $ \langle (X(T_f)- X(0))^2 \rangle $.  A linear fit yields $D_b(S)= 0.0052 S -0.0018$, valid for $S>S_0$, where $S_0$ is the breaking threshold slope. It is practically zero for $S<S_0$. 
(The range of $D_b$, as a function of $ S $,  is so small that it might be approximated by a constant: $D_b \approx 2 \times 10^{-4}$ m$^2$/s$^2$ ). 
 
%The parameter $D_b \approx D^* \theta(S-S_0)$, where $\theta$ is the Heaviside function. $D^*$ is approximately  
%$ 2 \times 10^{-4}$ m$^2$/s$^2$ and   $S_0$ is approximately  $0.3$.

\begin{figure}
	\centering
\includegraphics[width=5.1in,height=1.8in]{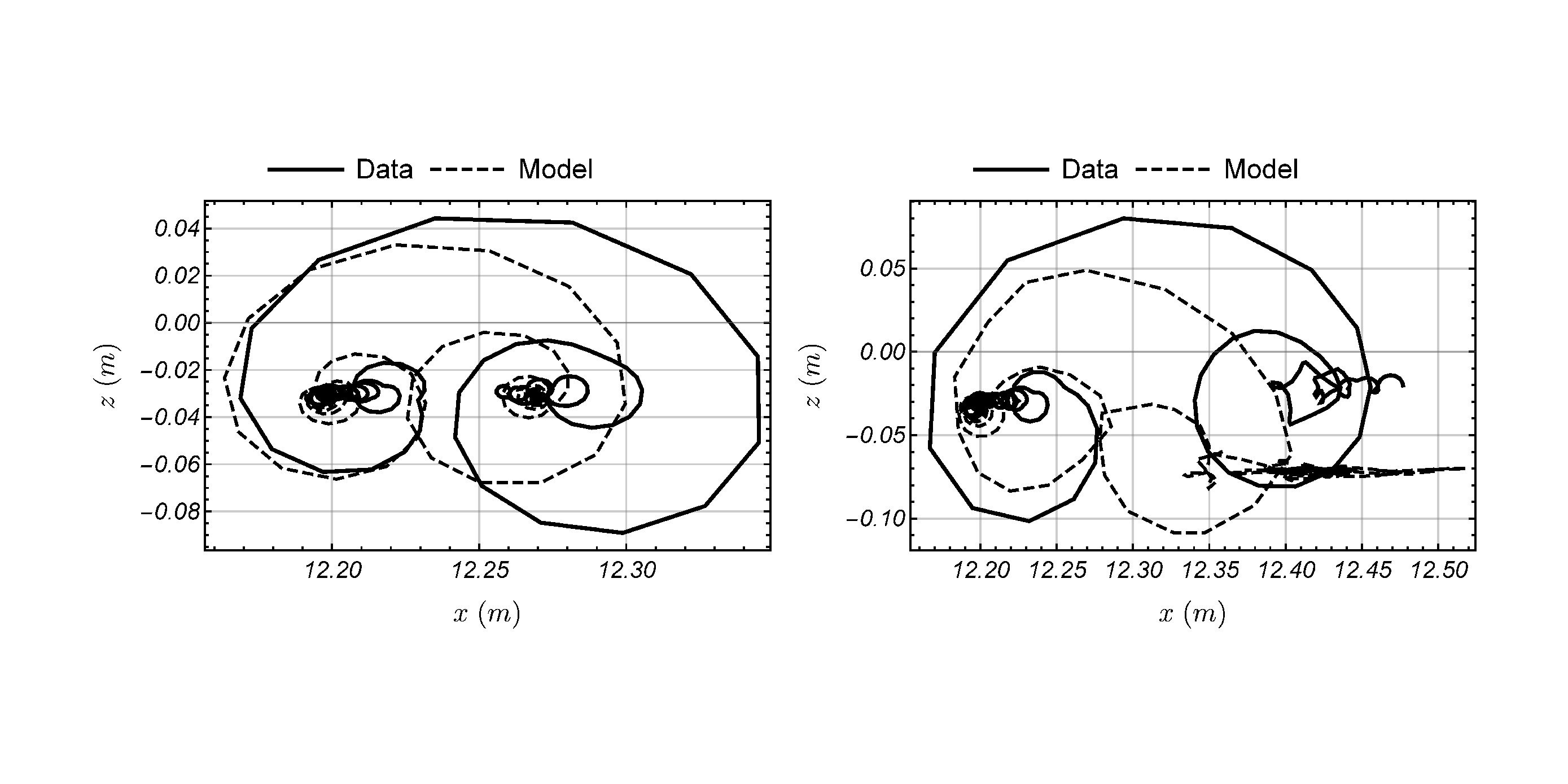}
	\caption{Comparison of  data (approximations to (\ref{transport})) and   the sub-surface model (\ref{ZR}) parcel paths, for  $ S=0.26 $ (left panel) and $ S = 0.4$ (right panel). $ \Zb(0) = (12.2,-0.3) $m in both cases. The dashed portion of the path corresponds to the motion of the parcel, close to the wave-breaking time.}
	\label{ZRfig}
\end{figure}

We will connote the  model given by  (\ref{ZR}) as the {\it sub-surface} mean transport model, in order to emphasize 
that this model tracks paths well for any $S$, including in the regime of $S$ leading to breaking waves, but does not apply to 
the near-surface (see Figure \ref{Fig_PointsTank}).  This is because it does not
take into account transport due to parcels that initiate in near-surface zone. As will be shown, 
the model captures the mean transport and dispersion, by virtue of the tuning of the parameters, for any $S$, below the gray zone.

\begin{figure}
\centering
\includegraphics[width=5in,height=4in]{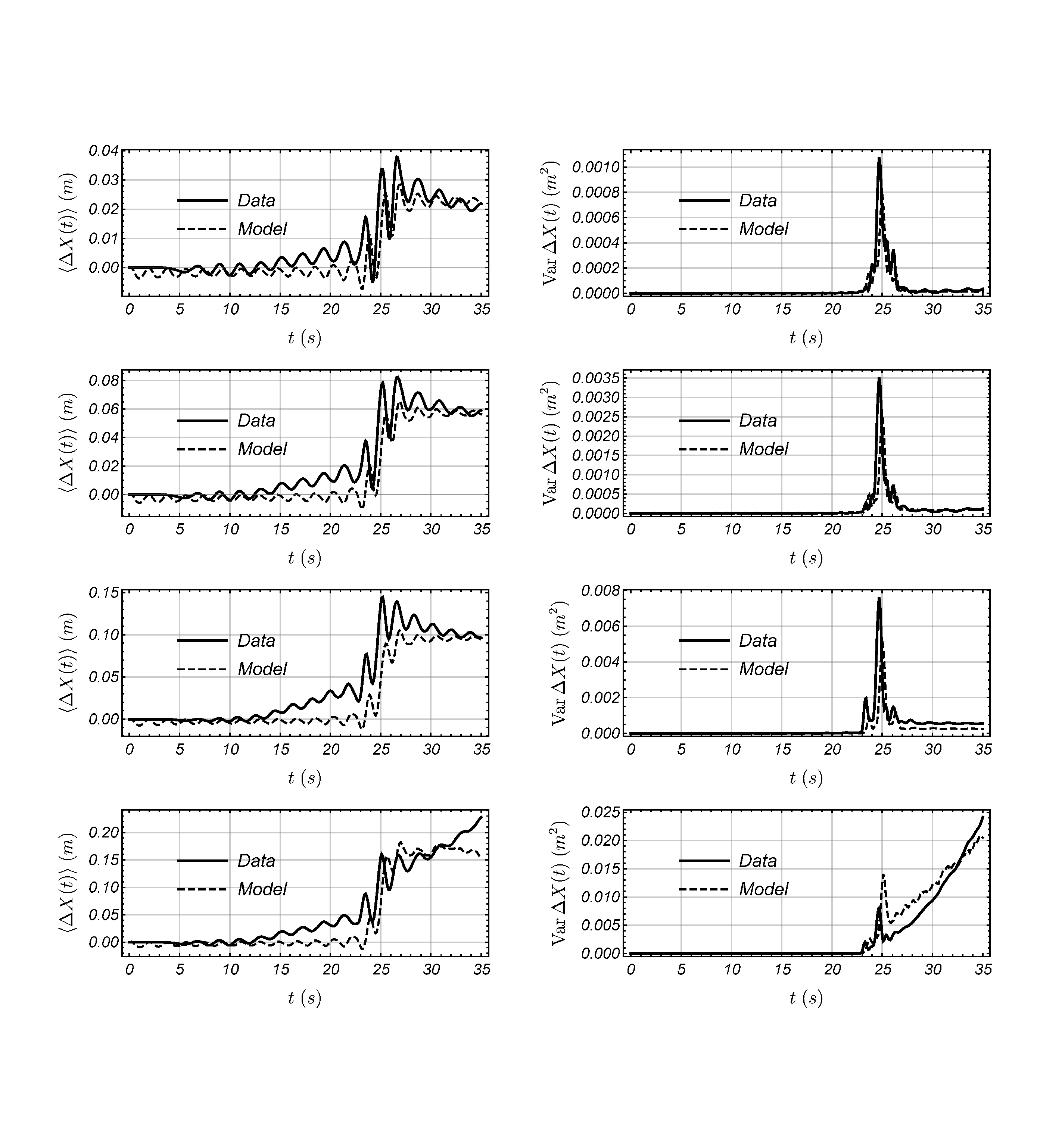}
\caption{ (Left) Ensemble mean of the horizontal displacement  $ \Delta X(t) = X(T_f) - X(0) $,  and (Right) its variance as a function of time.
From top to bottom, $S=  0.16, 0.256, 0.336, 0.416$, corresponding to 2 non-breaking cases and 2 breaking cases.  Averages are taken over all  trajectories starting at points marked in Figure \ref{Fig_PointsTank}.}
\label{tranmeanvar}
\end{figure}

Figure \ref{ZRfig} compares data to paths from a single realization of the sub-surface model in two contrasting cases:  a large amplitude, non breaking case,  $S=0.26$ in (a), and a breaking case, $S=0.4$ in (b).  The data and the model outcome are similar qualitatively, and close quantitatively.  Though not shown, for small $S$, the agreement between model and data is even better. Figure \ref{tranmeanvar} compares ensemble displacement and variance of the data
and model runs for $S=0.16, 0.256, 0.336, 0.416$, corresponding to 2 non-breaking cases and 2 breaking cases, respectively. The ensembles encompass data and model outcomes for all parcels with starting points indicated by dots  in Figure \ref{Fig_PointsTank}, but lying below the gray zone. 

\begin{figure}
	\centering
%(a)\includegraphics[width=5in,height=1.6in]{./TransportNoBreaking.pdf}
%(b)\includegraphics[width=5in,height=1.6in]{./DatavsModelAllS.pdf}
(a)\includegraphics[width=5in,height=1.6in]{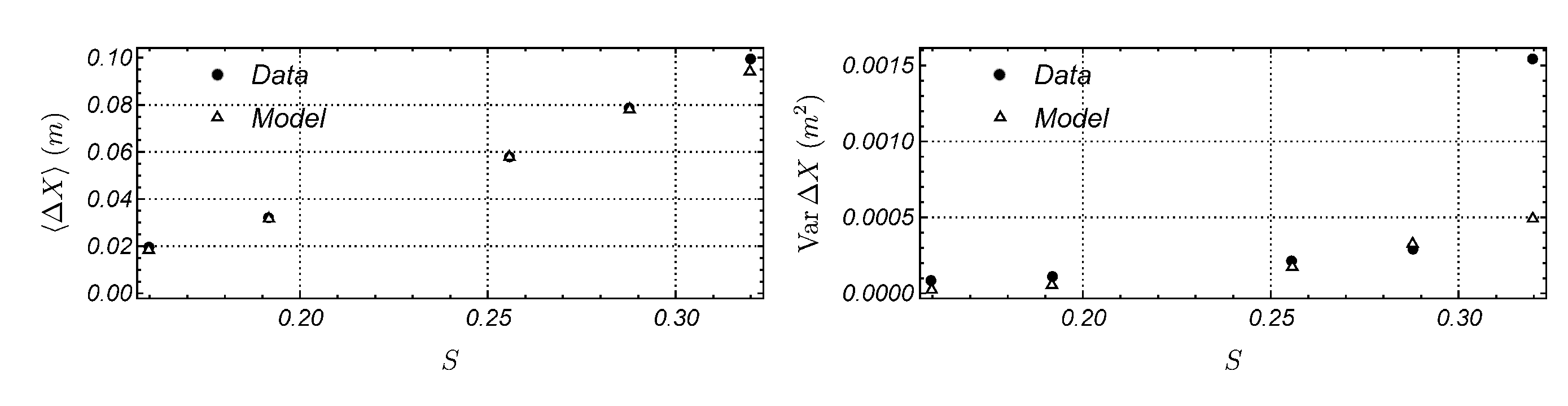}
(b)\includegraphics[width=5in,height=1.6in]{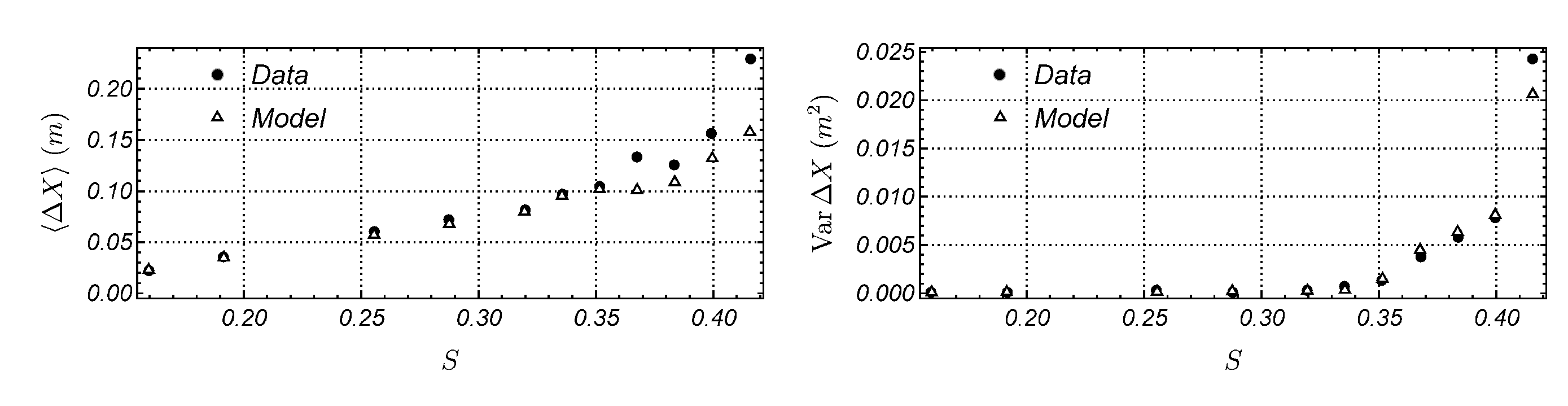}
	\caption{ Comparisons of the depth-averaged mean horizontal displacement  $ \Delta X(t) = X(T_f) - X(0) $ (left), and variance (right),  as a function of slope $S$. 
	Averages are taken over the trajectories starting in positions marked as black in Figure \ref{Fig_PointsTank}.}
	\label{transs}
\end{figure}

For non-breaking cases, $ S < S_0 = 0.336 $, the model is able to predict the statistics of transport for \textit{all} of the water column, including in the near-surface. Figure \ref{transs} displays comparisons of data and sub-surface model outcomes of the mean horizontal displacement and its variance, as a function of the slope $S$. In the top panels the experimental mean transport and variance is computed using {\it all} of the parcels, 
including parcels with initial conditions starting at the surface and the rest of the gray area. 
Conclusions from this comparison are  that the proposed model does a reasonably good job at predicting mean transport due to these particular progressive waves (and thus our analysis is borne out). By extension another  conclusion is that the mean transport is proportional to $S$ for large $S$ (breaking cases or otherwise), 
and it is numerically similar to the value predicted by a Stokes drift, and hence approximately proportional to $S^2$. Our analysis precludes
the possibility that there is a phase transition between breaking and non breaking transport, as alluded to in DPM17. In their analysis of the same data, mean transport for non-breaking cases is proportional to $S^2$ and changes to $S$ once wave breaking is present (see Figure 8 in DPM17).
 
In the lower two panels of Figure \ref{transs} we compare the mean transport and variance of the data and the model for the full range of $S$.
In these comparisons we exclude data and model outcomes from the near surface (gray region).

\section{Discussion and Summary}
\label{sec:summary}

%The transport is defined as the average of (\ref{transport}).
We analyzed  numerical simulations of the mean transport due to a transient,  progressive wave packet. The flow  was produced by a boundary-driven Navier-Stokes solver capable of resolving the free surface dynamics. The velocity boundary forcing has a parameter $S$, the wave slope, which controls the wave amplitude 
as well as its transient character.  
When $S$ is  sufficiently large the boundary forcing produces breaking waves. 

 Parcel paths were computed numerically using the velocity field obtained by the Navier-Stokes solver. 
The transport was analyzed by examining ensembles of the  parcel paths. In contrast 
to the steady, ideal non-breaking sinusoidal progressive wave case,  for which the mean transport is approximated by the Stokes drift regardless of 
wave amplitude,  transient wave packets can produce  more transport (along with  dispersion). This outcome obviously depends strongly on the transient nature of the wave itself. It was the case for the breaking and large non-breaking wave cases considered here.

The same data we analyzed here was examined in  DPM17. Their key conclusion is that mean transport is proportional to $S$ for breaking waves and to $S^2$ when waves do not. As a function of $S$  they suggest that  a phase transition in transport, from mean transport that is proportional to $S^2$ to transport proportional to $S$, coincides  with the occurrence of  wave breaking (see their Figure 8).
They conclude that   the sea surface `drift velocity due to breaking' is  proportional to $S-S_0$, where $S_0$ is a breaking  slope threshold.  
Their conclusion is  guided by approximate dimensional arguments, for breaking waves, which happen to correspond to large $S$ cases,  and the application of an analysis of parcel paths in \cite{pizzo17} (see Eqs. 224 and 4.1 in DPM17) of the John Equation for surface Lagrangian velocities (see \cite{johneq}).  However, the John equation predicts, trivially, that the transport is proportional to $S$, regardless of the value of $S$. It does not suggest, without further work,  how the mean transport is proportional to $S$, for large $S$, and to $S^2$ for small $S$.  Furthermore, well-posedness of the John equation precludes it applying to breaking waves.

  We  showed that, regardless of whether waves were breaking or not,  the parcel path evolution could be approximated, at least qualitatively,  by replacing
 ${\bf q}(\Zb,t)  \approx \ub^w(\Zb,t)$ in (\ref{transport}), where ${\bf q}$ is the Eulerian  velocity as approximated by the Navier-Stokes solver, and $\ub^w$ describes  the velocity of infinitesimal progressive  waves (the boundary forcing is generated by $\ub^w$). This approximation held for every parcel, and for all  values of $S$, so long as there was no free surface breaking.   For small $S$ and for parcels far from the sea surface the above approximation is even quantitatively close.  However, for $S>S_0$, in the range of $S$ in which wave breaking occurs, 
  the approximation still held but not for parcels in the near-surface layer.

  When $S$ is small, the mean transport is of order  $S^2$. A transport of  order $S^2$ would be expected of non-breaking steady progressive waves, regardless of their amplitude. This is because the time average of  $\ub^w$ is vanishingly small, and thus the mean Lagrangian 
 transport is approximated by the second order correction, the Stokes drift velocity. 
 For large $S$ the mean transport of the transient waves was instead proportional to $S$, and this is  because the time average of
 $\ub^w(\Zb,t)$ in (\ref{Z2}) does not vanish for these waves.
   The  mean transport varies smoothly with $S$: for small $S$ the waves exhibits transient behavior, and it increases with $S$. 
  At the same time, with increasing $S$,  the average of  $\ub^w(\Zb,t)$  transitions smoothly in value from very small, to values proportional to $S$. There is no phase transition in the mean transport, over the range of $S$.

   For large $S$ and for parcels starting at or very near to the free surface, the estimate $\dot{ \Zb} \approx \ub^w(\Zb,t)$ overestimates the extent of the orbital paths, and does not capture the resulting dispersion of the ensemble of paths. Apparently 
 boundary layer effects and the generation of vorticity (not shown in this work but confirmed in the simulations) are but two manifestations of the difference between the Navier-Stokes 
 velocity ${\bf q}$ there and the approximation $\ub^w$. The parcel paths, near the free surface, for $S>S_0$ cases, can be tortuous, dispersive and thus cannot produce orbits as large as those given by $\dot {\bf Z} \approx \ub^w$. This observation is consistent  with established notions on  wave breaking phenomena that  describe aspects of the wave energy and momentum transfer to the mean flow, and the resulting generation of shearing and vorticity (see \cite{craigbanner}. See also the review in \cite{Melville96}).

%Our analysis is straightforward: for small $S$ the mean $\ub^w(\Zb(t),t)$ is vanishingly small, because the waves have a very weak transient evolution, and the second order correction captures the {\it mean} transport. This order $S^2$ mean transport is approximately, the monochromatic case, the Stokes drift. For large $S$, however, the average of   velocity is not small and thus the mean transport is proportional to the velocity  itself (which is   proportional to $S$).

 Better quantitative comparisons of the mean transport were obtained when we incorporated  an empirical dissipation term into the velocity: we replaced $\ub^w(\Zb,t)$ by $\ub_d^w(\Zb,t)$. See (\ref{paddled}). The empirical dissipation term 
 takes every $n$ component in the spectrum of the linearized irrotational velocity and multiplies it by $\exp [-\beta k^2_n t]$, $\beta>0$ a fitted constant. This is, arguably, a simplistic model for dissipation for  Navier-Stokes solutions, especially those involving wave breaking, however, it is telling that standard linear damping is all that is necessary to get good mean transport
 results. The damping term, however, does not capture the dispersion evidenced in the data for large $S$ cases.
 When $S$ is large vorticity appears very close to  free surface, and intermittent dispersion is evidenced in the data. When $S>S_0$, vorticity and  dispersion appear and linger, even after the passage of the breaking wave. Hence,
the breaking waves introduces  non-trivial parcel dispersion.  There is an exchange of momentum between waves and the mean flow and 
the ${\bf q}(\Zb,t)  \approx \ub_d^w(\Zb,t)$ remains qualitatively acceptable, so long as it is not used to capture the paths of parcels that begin close to the free surface or on the free surface. The irrotational approximation $ \ub^w(\Zb,t)$ makes the Lagrangian  velocity overshoot and further, the approximation does not take into account dispersion.

These observations form the basis for a simple parametric stochastic mean transport model, applicable for any $S$. The parcel path velocity is modeled by a combination of the damped velocity   $\ub_d^w(\Zb,t)$ and a stochastic term that  generates parcel dispersion and introduces
irrotationality in the resulting Eulerian velocity. The expectation of the model leads to a simple model of the mean transport that agrees with the mean transport in the data. Specifically,  the model (see (\ref{ZR}))  compares favorably with data  in estimating the mean transport  for small $S$ and for large $S$, even when breaking occurs (see Figures \ref{tranmeanvar}-\ref{transs}).  The model  yields good estimates of mean transport  at any depth, including at the surface, so long as breaking is excluded. However, it yields good results even when breaking takes place, outside of the wave layer. 
 The stochastic aspect of the model is  crude but suggests that a stochastic  modeling approach might be useful in capturing parametrically the effect of wave breaking on transport and dispersion, albeit in the mean.
 Agreement between the data and the model was achieved by tuning two model parameters.  The parameters in the model were easy to tune to the data, however, it is not clear how well the model and its tuning would fare in an oceanic setting. Nevertheless, the utility of the model lies mostly in its ability to explain the details of the transport  in the whole `tank'  and for the whole range of $S$ (the whole tank for $S<S_0$).

  In oceanic situations transient waves are common, and thus understanding under what circumstances  transients produce significant transport is worth further consideration. Because the circumstances depend on the time spans in question, it is perhaps best to study transient transport as it relates to specific phenomena.
   For example, one might want to determine how Langmuir turbulence and oceanic processes that depend critically on wave-generated transport are affected by waves with transient  behavior. The same sort of question could apply  to the transport and dispersion of tracers, such as pollutants and of phytoplankton, and to air-sea exchanges.

%
% By means of an exemplary flow our analysis suggests that  transient wave transport and dispersion is of significance, at the wave scale and beyond, and worthy of a more systematic and comprehensive analysis.  Proposals on  how Langmuir turbulence, for example,  is modified to include wave breaking events appear in  \citealt{SMM07}, \cite{RMMB}. Langmuir turbulence may be an important problem that might 
% frame a research agenda into the impacts of  wave-induced transport and dispersion under transient conditions.

Acknowledgements: 
We thank L. Deike and K. Melville for sharing their data and for helpful discussions. We also thank James C. McWilliams for stimulating discussions and for suggesting improvements to the manuscript. 
This work was supported through a grant by the National Science Foundation, NSF OCE1434198.
JMR wishes to thank the Kavli Institute of Theoretical Physics at the University of California, Santa Barbara. The KITP is supported in part by the National Science Foundation under Grant No. NSF PHY-1748958.

\bibliographystyle{jfm}
\bibliography{swwe}

\end{document}